\documentclass[journal=jpcafh,manuscript=article,layout=twocolumn]{achemso}

\usepackage[english]{babel}
\usepackage[utf8]{inputenc}
\usepackage{amsmath}
\usepackage{graphicx}
\usepackage{color}
\usepackage{subfigure}
\usepackage{multirow}     
\usepackage{pifont}
\usepackage{soul}

\usepackage{dcolumn}% Align table columns on decimal point
\usepackage{bm}% bold math
\newcommand{\good}{{\ding{52}}}
\newcommand{\bad}{{\ding{56}}}


\title{Combining Pseudopotential and All Electron Density Functional Theory for the Efficient Calculation of Core Spectra using a Multiresolution Approach}

\title{Combining Pseudopotential and All Electron Density Functional Theory for the Efficient Calculation of Core Spectra using a Multiresolution Approach}

\author{Laura~E.~Ratcliff}
\email{laura.ratcliff08@imperial.ac.uk}
\affiliation{Department of Materials, Imperial College London, London SW7 2AZ, UK}
\alsoaffiliation{Argonne Leadership Computing Facility, Argonne National Laboratory, Illinois 60439, USA}

\author{W.~Scott~Thornton}
\affiliation{Stony Brook University, Stony Brook, New York 11794, USA}

\author{\'Alvaro~V\'azquez~Mayagoitia}
\affiliation{Computational Science Division, Argonne National Laboratory, Illinois 60439, USA}
\alsoaffiliation{Argonne Leadership Computing Facility, Argonne National Laboratory, Illinois 60439, USA}

\author{Nichols~A.~Romero}
\affiliation{Computational Science Division, Argonne National Laboratory, Illinois 60439, USA}
\alsoaffiliation{Argonne Leadership Computing Facility, Argonne National Laboratory, Illinois 60439, USA}

\date{\today}

\begin{document}

%\begin{tocentry}
%\includegraphics[]{cover_drawing.png}
%\end{tocentry}

\begin{abstract}
Broadly speaking, the calculation of core spectra such as electron energy loss spectra (EELS) at the level of density functional theory (DFT) usually relies one of two approaches: conceptually more complex but computationally efficient projector augmented wave based approaches, or more straightforward but computationally more intensive all electron (AE) based approaches.
In this work we present an alternative method, which aims to find a middle ground between the two.  Specifically, we have implemented an approach in the multi-wavelet \textsc{madness} molecular DFT code which permits a combination of atoms treated at the AE and pseudopotential (PSP) level.  Atoms for which one wishes to calculate the core edges are thus treated at an AE level, while the remainder can be treated at the PSP level.  
This is made possible thanks to the multiresolution approach of \textsc{madness}, which permits accurate and efficient calculations at both the AE and PSP level.  Through examples of a small molecule and a carbon nanotube we demonstrate the potential applications of our approach.
\end{abstract}

%\maketitle

\section{Introduction}

\textsc{madness} (Multiresolution ADaptive Numerical Environment for Scientific Simulation)~\cite{madness_github,Harrison2016} is a general purpose numerical framework which combines a multiresolution approach with a parallel programming environment designed for petascale performance.~\cite{VAZQUEZMAYAGOITIA20143}  The use of an adaptive multiresolution approach allows integral and differential equations in many dimensions to be solved with guaranteed precision.  Furthermore, the code has been structured so that developers can focus on the high level implementation of new functionalities, without needing detailed knowledge of the low level technicalities. These features have facilitated the development of various scientific applications using \textsc{madness}, including a molecular density functional theory (DFT) code~\cite{Harrison2003,Harrison2004,Yanai2004}, and many other applications spanning a range of fields, many of which are concerned with quantum chemistry~\cite{Bischoff2012,Bischoff2013,Bischoff2017,Fann2009,FossoTande2013,Kottman2015,Reuter2012,Reuter2012a,Sekino2008,Sullivan2013,Vence2012,Yanai2004a,Yanai2004b}.

The molecular DFT code (hereafter referred to as moldft) provides a setup for the precise treatment of electronic systems with an excellent cost to accuracy ratio.  Nonetheless, the cost of treating heavy atoms -- or large molecules containing many light atoms -- remains high, inhibiting the ability to study technologically interesting systems such as those containing transition metals.  The most popular way to alleviate this problem is the replacement of the exact all electron (AE) atomic potential with a smoother pseudopotential (PSP), which has the dual advantage of reducing the number of electrons requiring explicit treatment (as the core electrons are incorporated implicitly), and reducing the number of basis functions needed to represent the wavefunctions close to the nuclei, thanks to increased smoothness.  This enables heavy atoms to be treated with a much lower computational cost and would extend the applicability of moldft.

Beyond these advantages, the multiresolution approach of \textsc{madness} also facilitates the implementation of a \emph{mixed} representation, namely the treatment of only select atoms with PSPs, with others retaining the full AE potential.  While this is also possible in Gaussian basis sets using effective core potentials (ECP), a multiresolution approach permits a similar method using a \emph{systematic} basis, since the automatic refinement ensures that both full and pseudo-atoms are treated at a resolution which is high enough to maintain accuracy without loss of efficiency.

One application where this is of particular use is core level spectroscopy, e.g.\ electron energy loss spectra (EELS), specifically energy loss near edge structure (ELNES), which is the region of the spectrum immediately after the edge onset.  ELNES is able to probe the local electronic structure of a sample and is thus an invaluable experimental technique. Theoretical calculations (often based on DFT) are required to help interpret experimental spectra.  The development and application of methods for simulating ELNES is therefore an active area of research~\cite{Ikeno2017}.
However, one is often only interested in calculating spectra for a subset of a system, e.g.\ a group of atoms or single atomic species within a molecule, or a molecule in an environment.
In such cases, core states need only be explicitly calculated for the atoms of interest, which can easily be achieved by treating only these atoms at the AE level.

In this paper we first outline the key concepts behind the \textsc{madness} molecular DFT code, before briefly discussing the implementation of a mixed AE/PSP approach.  We then present the method used to calculate ELNES, including the calculation of the virtual Kohn Sham (KS) eigenstates.  Finally, we validate the approach through examples of two  systems which benefit from a mixed PSP/AE representation.

\section{Theory}

\subsection{Molecular DFT with MADNESS}

Both the \textsc{madness} code as a whole and the moldft code in particular have already been described elsewhere~\cite{Harrison2016,Harrison2003,Harrison2004,Yanai2004}, and so here we give only an outline.
One of the central components of \textsc{madness} is the use of a disjoint ``multi-wavelet'' basis set, which is constructed from a set of (shifted and scaled) Legendre polynomials represented in a non-uniform grid.  The grid, therefore the basis, is dynamically adapted to give higher resolution where needed (e.g.\ close to the atoms where the KS wavefunctions are more rapidly varying), giving rise to a computationally efficient yet highly accurate multiresolution analysis (MRA).  Indeed, each KS orbital has its own individual adaptively refined representation.  The exploitation of MRA techniques allows the code to reach a finite arbitrary precision, while relieving the user of the need to manually converge the basis set.

In contrast to other DFT codes, the central eigenvalue equation to be solved is recast from a differential equation of the form
\begin{equation}
\left(-\frac{1}{2} \nabla^2+V\left(\mathbf{x}\right)\right)\psi\left(\mathbf{x}\right)=E\psi\left(\mathbf{x}\right) \; ,
\end{equation}
to an integral equation of the form
\begin{equation}\label{eq:int}
\psi\left(\mathbf{x}\right)=-2\int \mathrm{d}\mathbf{x}\left(-\nabla^2-2E\right)^{-1}V\left(\mathbf{x}\right)\psi\left(\mathbf{x}\right) \; .
\end{equation}
This form is well-suited to the \textsc{madness} framework and has the advantage of being able to be solved iteratively without the need for a preconditioner. 
In practice, an initial guess for the KS wavefunctions is first generated by projecting an atomic orbital basis, typically 6-31G, into the multiwavelet basis.
This approach takes advantage of the underlying \textsc{madness} numerical and parallel runtime to efficiently solve the KS equations to a guaranteed precision, while requiring only minimal input from the user concerning the basis set or parallel setup.

A wide range of LDA, GGA and hybrid functionals are available in moldft through the \textsc{LibXC} library~\cite{Marques2012}.
Although we have not exploited the capability in this work, in addition to standard canonical orbitals, moldft has the option to use localized molecular orbitals~\cite{Yanai2004a}, where the Pipek-Mezy scheme~\cite{Pipek1989} is used to localize the orbitals.  This  gives rise to a quasi-linear scaling, reducing the computational cost of treating larger systems.

\subsection{Pseudopotentials in MADNESS}

Despite the advantages of an MRA approach for DFT calculations, the computational cost of treating heavy atoms or indeed of large systems containing light elements is high, due in part to the explicit treatment of core electrons. 
The applicability of moldft could therefore be extended to new materials by implementing PSPs.
Although most commonly used in periodic plane wave (PW) DFT codes, PSPs are also effectively employed in combination with other basis sets.  
Given such extensive use for DFT calculations, it is of no surprise that PSPs exist in a number of varieties, generally categorized as ``hard'' or ``soft'', depending on the smoothness of the pseudized-wavefunctions.
Indeed, the development of ever cheaper and more accurate PSPs continues to be an active area of research. 
It is also worth mentioning the projector augmented wave (PAW) approach~\cite{Blochl1994}, an alternative approach to more traditional PSPs, which aims to reproduce the correct AE behaviour near the nuclei, while still avoiding the explicit treatment of core states.

We have chosen to implement norm-conserving HGH-GTH~\cite{Goedecker1996,Hartwigsen1998} PSPs in moldft, as they are available for the majority of elements and have demonstrated a consistently high accuracy and transferability (see e.g.\ Refs.~\cite{Lejaeghere2016,Willand2013}).
Furthermore, they have already proven to work well in the  wavelet-based BigDFT code~\cite{Genovese2008}, for calculations in both open and periodic boundary conditions. This also provides a means of validating our implementation.

The implementation in \textsc{madness} was relatively straightforward, since the underlying machinery used to solve the KS equations remains the same, only the definition of the atomic potential need be modified.
In the first instance we have not implemented either relativistic effects or non-linear core corrections, which have demonstrated an accuracy of similar quality to AE calculations~\cite{Willand2013}.  In the future this might easily be extended.

\subsubsection{Mixed AE/PSP Calculations}

As discussed, an adaptive multi-wavelet approach is also highly suitable for a mixed AE/PSP representation.
Both the implementation and application of such an approach is straightforward.  Consider two opposing scenarios: calculations using PWs and those using Gaussian-type basis sets.  For the former, aside from the prohibitive cost of AE calculations in PW basis sets, the delocalized nature of the basis functions also prevents them from being spatially varied to be more or less dense around different atoms.  Any mixed AE/PSP approach would therefore also necessitate a mixed basis set approach, e.g.\ combining PWs with a localized basis set~\cite{Louie1979,VandeVondele2005}.

In the case of Gaussian basis sets, the number of functions associated with each atom could be directly modified depending on the level of theory used.  Such a mixed representation has previously been implemented using Gaussian basis sets and used to assess the accuracy of individual PSPs for molecular properties including binding energies and vibrational properties~\cite{Briley1998}.  This approach has been used to treat select groups of atoms at the AE level, for example H atoms in molecules or clusters for the calculation of Raman spectra~\cite{Briley1998,Jackson2001}, only the molecule for the study of adsorption on a surface~\cite{Porezag2000} and for M\"{o}ssbauer-active Sn atoms for the calculation of M\"{o}ssbauer spectra in chalcogenide glasses~\cite{Jackson2002}.

While a mixed approach can therefore clearly be employed using Gaussian basis sets, this nonetheless increases the burden on the user to ensure that the simulation remains both accurate and computationally efficient by tuning the basis for each atom according to chemical intuition. Furthermore, the use of the same underlying approach to e.g.\ the calculation of Poisson's equation for PSP and AE approaches did not guarantee a reduction in computational cost for mixed AE/PSP compared to pure AE calculations~\cite{Briley1998}.
The key advantage of a multi-wavelet approach is therefore that the same high accuracy and computational efficiency is achieved for the mixed mode as for pure AE and PSP approaches.  Crucially, this does not require any additional fine-tuning of parameters -- to perform a calculation in mixed mode, the only thing the user must do is specify which atoms are to be treated at the PSP level.

\subsection{Calculating ELNES}

One way of maximizing the information that can be extracted from DFT simulations is by calculating experimental quantities like spectra. For example, core spectra such as ELNES can be used to extract information concerning the chemical bonding environment, valence state and nearest neighbour distances.  The simulation of such spectra is invaluable both for understanding and interpreting experimental results and predicting and guiding future experiments on new materials. To give an example, a combined theoretical and experimental approach allowed the identification of individual fullerene molecules encapsulated in a carbon nanotube~\cite{Tizei2014}.  Here we summarize the different approaches to calculating ELNES within AE and PSP DFT calculations.  For a more thorough discussion of the calculation of ELNES see e.g.\ Refs.~\cite{Pickard1997,Ikeno2017}.

The most straightforward method of calculating ELNES within DFT (beyond the simple site- and angular-momentum-projected density of states approach) is via Fermi's golden rule in conjunction with the dipole approximation.  In this formalism the imaginary part of the dielectric function, $\varepsilon_2$, in atomic units, is given by
\begin{equation} \label{eq:eps2}
\varepsilon_2 \left( \omega \right) = \frac{1}{\Omega} \sum_{c, v} {\lvert \langle \psi_v \lvert \mathbf{q} \cdot \mathbf{r} \rvert \psi_c \rangle \rvert}^2 \delta \left( E_v - E_c - \omega \right) \; ,
\end{equation}
where $\omega$ is the transition energy, $\Omega$ is the volume of the unit cell, $\mathbf{q}$ is the momentum transfer, $\mathbf{r}$ is the position operator and $\psi_c$ ($\psi_v$) is a core (virtual) state with associated energy $E_c$ ($E_v$).  In practice, the $\delta$-function is usually replaced by a Gaussian or Lorentzian function to simulate broadening effects or lifetime of the transition. 

If one assumes that the excitation of a core electron to a virtual state is ``sudden'', i.e.\ the virtual states are unaffected by creation of an instantaneous core hole (and neglecting relativistic effects), one can directly calculate the matrix elements of Eq.~\ref{eq:eps2} between core and virtual KS eigenstates from a ground state calculation.
This provides a first approximation, however more quantitative comparison with experiment generally requires the inclusion of core hole effects.  Different approaches may be used, the most basic being the $Z+1$ approximation, wherein the excited atom is replaced by an atom with atomic number one higher than its actual atomic number.  This has met with mixed success, see e.g.\ Ref.~\cite{Duscher2001}. Alternatively, a PSP may be generated with a missing core electron~\cite{Gao2009}, while for AE calculations a constraint may be applied to maintain a core hole.  Irrespective of the approach used, a separate calculation is needed for each atom one wishes to excite.

For AE-based DFT approaches, the calculation of matrix elements between core and virtual states is straightforward.
For PSP approaches where there are no explicit core states, one could instead use core states originating from an isolated atom.  However the calculated (valence and) virtual states are pseudo-wavefunctions and therefore only match the true wavefunctions outside of the core region.  This can have a noticeable impact on the accuracy of the matrix elements.
If one assumes that the core wavefunctions are themselves unaffected by their environment, one could nonetheless calculate accurate matrix elements if the correct behaviour of the virtual states could be recovered in the core regions.  This can be achieved using the PAW approach, which is able to recover the AE valence and virtual wavefunctions, resulting in matrix elements which are comparable in accuracy to AE approaches while requiring significantly less computational effort.
In contrast, ELNES calculations using standard PSPs (i.e.\ without PAW), generally do not give accurate values for the matrix elements~\cite{Gao2009,Pickard1997}.

PAW has been successfully employed for ELNES calculations using a PW basis set~\cite{Mazevet2010,Gao2009,Pickard1997} for systems which are larger than could be accessed using AE approaches, where PAW is either used in place of PSPs or as a correction to the ELNES matrix elements following a PSP calculation.
Indeed, more than 1000 atoms have been treated using a PAW approach within the context of linear scaling DFT calculations with a psinc basis~\cite{Tait2016}. 
This allows the treatment of more complex materials, while also permitting supercell calculations which are large enough to minimize interactions between core holes in periodic images.

The absolute energies of the core levels for a given atomic species are affected by their local chemical environment.  The magnitude of such variations, i.e.\ the chemical shift, depends on the species in question -- e.g.\ for the C~1s state this can be as much as 12~eV~\cite{Gelius1970}.  Except in trivial cases where all atoms of a given species within a molecule have the same chemical environment, such shifts must therefore be taken into account in order to correctly combine spectra originating from different atoms of the same species within a system. This cannot be achieved by manually aligning theoretical and experimental spectra, instead one must calculate the absolute energy onsets.

A first approximation to calculating such energies would be to take the difference between the corresponding KS eigenvalues.  This relies on having access to the energies of the core states, so that such an approach is only applicable within an AE calculation.  A better approach would be to also include core hole effects in the calculated energies, by taking the difference between the total energy of the excited (core hole) and ground state calculations, which relies on the explicit inclusion of core states.  However, it is possible to estimate the absolute energy onsets from PAW calculations by also calculating the effect of core holes on isolated atom calculations~\cite{Mizoguchi2009}.

The mixed AE/PSP approach provides an interesting compromise between the AE and PAW approaches for calculating ELNES.  In cases where one is only interested in calculating excitations for few atoms of interest, the computational cost is significantly reduced compared to a pure AE approach, while the direct access to select core states allows for the accurate and straightforward calculation of transition matrix elements and energies.

Since the goal of this paper is a demonstration of the benefits of a mixed PSP/AE representation within a multiresolution approach, rather than the explicit comparison of calculated ELNES spectra with experiment, the implementation of core hole effects is left as a future extension, where the constrained approach mentioned above should be used -- note that it would already be possible to employ the $Z+1$ approach.  Since we have not included core hole effects we calculate the transition energies by taking the difference between KS eigenvalues.

\subsubsection{Calculation of KS States}

The KS core and virtual (unoccupied) states are a key ingredient of Eq.~\ref{eq:eps2}.  In moldft, a given number of virtual states
can be calculated alongside the occupied KS states.
We note that the approach is not designed to allow access to unbound states -- with positive energies the kernel of Eq.~\ref{eq:int} ($-(\nabla^2-2E)^{-1}$) diverges -- while in any case such states might strongly depend on the simulation cell size.
We therefore avoid calculating such eigenstates in this work.
However, there is also another subtlety relating to the initial ordering of the (bound) virtual states.  This is similar to a situation which arises in the context of the linear-scaling DFT code \textsc{onetep}~\cite{Skylaris2005}, wherein the virtual states are represented in a localized orbital basis set, which is itself represented in an underlying systematic basis set.  Starting from an initial atomic orbital guess, these localized orbitals are optimized to  represent a set of low lying virtual states.   However, the virtual states can be incorrectly ordered in the initial basis, so that some high energy virtual states are selected in favour of states which would be lower in energy in an optimized (i.e.\ more complete) localized orbital basis, resulting in some ``missing'' virtual states~\cite{Ratcliff2011}.

While we are not applying any localization to the KS states, the initial guess is nonetheless generated from a localized atomic basis set, so that the same energy ordering problems can occur.   Fortunately, the same solution can also be used: a larger number of virtual states than required must initially be requested to allow the virtual states to attain the correct energy ordering.  After this initial stage the higher energy KS states are eliminated and the calculation proceeds with the actual number of states required. 
In order to reduce the user effort, this process has been semi-automated so that the code will gradually reduce the number of calculated virtual states, however the user must still pay careful attention to ensure no virtual states have been neglected.

There are also some subtleties regarding core states.  When multiple atoms are treated as AE, it is possible for mixing to occur between core wavefunctions.  For standard DFT calculations this does not pose a problem, however when we are interested in probing excitations originating on specific atoms this is problematic.  One way to minimize this mixing is by explicitly localizing the core states, with the aim of ensuring that the core states remain associated with a single atom.
In some cases it might be necessary to impose a more strict localization criterion, however in the following examples this was not needed.  Furthermore for core hole calculations one could always treat only a single atom at the AE level, avoiding such problems.

\section{Results}

We present in the following benchmark calculations demonstrating the accuracy of our approach for cysteine and a single walled carbon nanotube (SWCNT) by comparing the PSP, mixed and AE approaches with results from other codes employing different basis sets and approaches.
We emphasize here that the goal is not to present a new approach for the calculation of core spectra, but to demonstrate the potential advantages of using a multiresolution approach for calculating such quantities in a straightforward and computationally inexpensive manner. 
Our motivation behind calculating ELNES is  to show an example of a type of calculation where the use of a mixed AE/PSP scheme within a multiresolution approach is advantageous.
As such, we have not presented any comparisons with experiment, as many such comparisons, including the impact of whether of not core hole effects are incorporated, can be found elsewhere; see for example a recent review article on the subject~\cite{Ikeno2017}.

\subsection{Computational Details}

All ELNES spectra were generated for an isotropic average of $\mathbf{q}$. 
For \textsc{madness}, \textsc{BigDFT}~\cite{Genovese2008} and \textsc{NWChem}~\cite{Valiev2010} we used explicit free boundary conditions, while for \textsc{castep}~\cite{Clark2005} we used large cubic supercells with sides of 30~\AA\ to minimize interactions between periodic images.
For \textsc{madness} and \textsc{BigDFT} we used the same PSP parameters, ensuring that the employed PSPs were generated with the correct functional, while for \textsc{castep} we used the on-the-fly PSP generator using default parameters.
Following a PSP calculation in \textsc{castep}, PAW is used to correct the ELNES matrix elements, as described in Refs.~\cite{Pickard1997,Gao2009}.
For cysteine we used the LDA exchange correlation functional~\cite{Ceperley1980}, while for the SWCNT we used the PBE functional~\cite{Perdew1996}.

\textsc{madness} calculations were initially converged using a threshold of $10^{-4}$ and wavelet order $k=6$, following which $10^{-6}$ and $k=8$ were used to achieve a well converged result.
For \textsc{NWChem} we used aug-cc-pV$n$Z basis sets with varying $n$, which we abbreviate to aV$n$Z in the following.
For \textsc{BigDFT} we employed a small wavelet grid spacing of 0.08~\AA\ and coarse and fine radius multipliers of 12 and 15 respectively to obtain well converged energies.
For \textsc{castep} we used PW cutoffs of 1000~eV and 600~eV for cysteine and the SWCNT respectively.
In all codes we calculated the five lowest energy virtual states for cysteine and the fourteen lowest energy virtual states for the SWCNT.

\subsection{Cysteine}

\begin{figure}[!h]
\centering
{\includegraphics[width=0.25\textwidth]{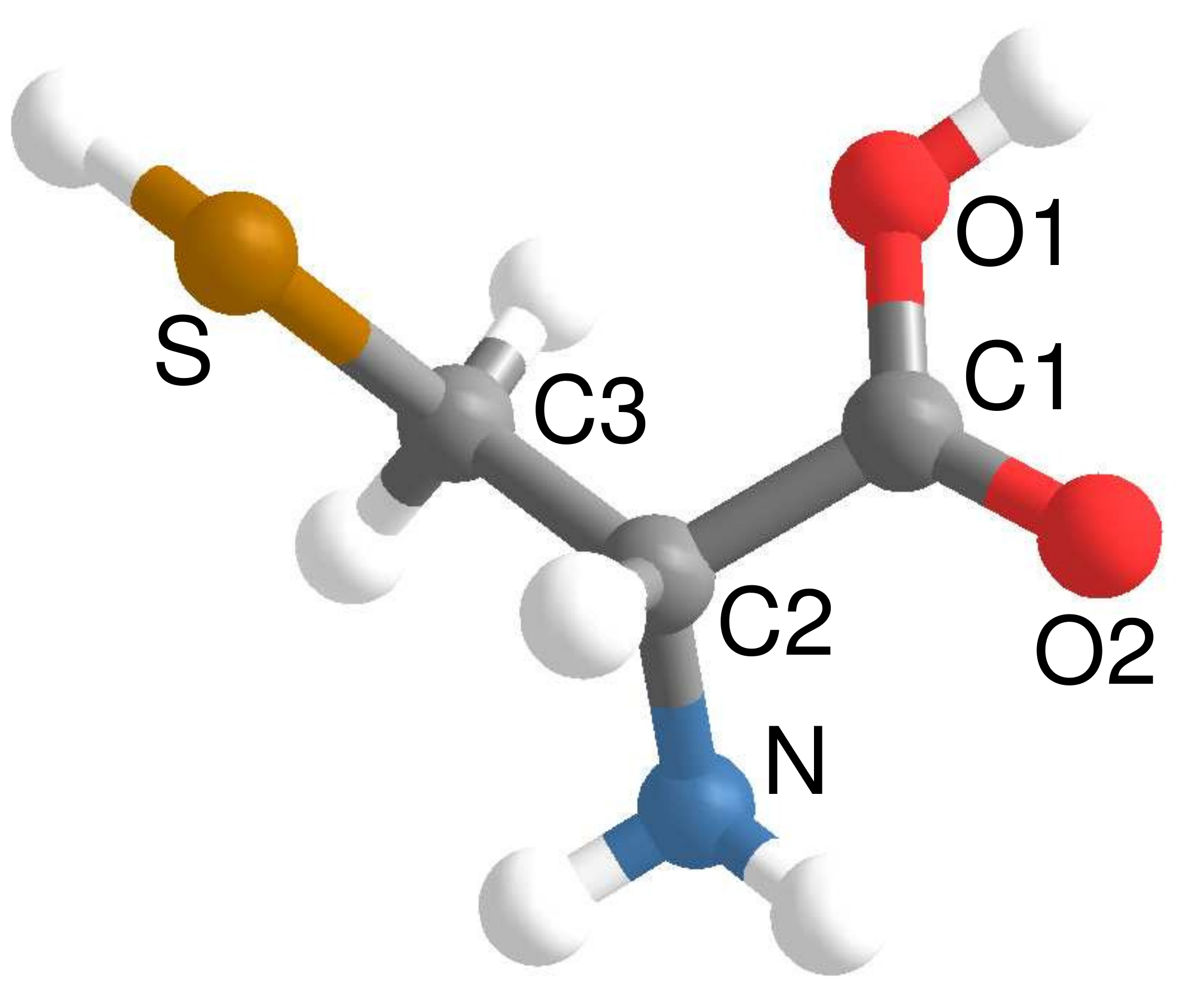}}
\caption{Atomic structure of the cysteine molecule, with H atoms in white, C in grey, N in blue, O in red and S in gold. The atom labels refer to those used in Table~\ref{tab:cysteine_core} and Fig.~\ref{fig:cysteine_eels_atoms}.}
\label{fig:cysteine_mol}
\end{figure}

We first take the amino acid cysteine, for which the atomic structure is depicted in Fig.~\ref{fig:cysteine_mol}.  We use this example to verify the correctness of the PSP implementation and assess the mixed AE/PSP scheme.  In order to put the results into context we also compare with other codes using different basis sets.

We first compare select eigenvalues close to the HOMO (highest occupied molecular orbital) and the LUMO (lowest unoccupied molecular orbital) -- results are shown in Table~\ref{tab:cysteine_eigen}.
In the first instance, we compare the AE and PSP approaches in moldft with the PSP implementation in \textsc{BigDFT}, which, as previously mentioned, employs the same type of PSPs. 
The PSP results from the two codes are in excellent agreement, with differences of at most a few tenths of a meV, confirming the correctness of our PSP implementation.
Comparing the AE and PSP values, as expected the differences are more significant, but nonetheless remain small, of the order of 10~meV.  Such a high level of agreement is more than sufficient given typical smearing applied when calculating ELNES.

To put the results in context, we have also calculated the energies using the \textsc{castep} PW code and with various Gaussian basis sets using \textsc{NWChem}, as well as the Stuttgart relativistic large core ECP~\cite{Bergner1993} in \textsc{NWChem}.  For the latter, ECP were used for all elements except H, for which the aVTZ basis was used.  This approach gives the largest errors compared to the \textsc{madness} AE results, which might be attributed to the fact that they have been developed for Hartree Fock calculations rather than DFT.
The next biggest discrepancies are for \textsc{castep}.  
This is unsurprising given the greater differences in computational setup, including the use of a periodic supercell and different PSP type.  However the differences are  small -- less than 50~meV.  The error due to using the smallest Gaussian basis set (aVDZ) is of a similar magnitude, while for the largest Gaussian basis set (av5Z) the difference with respect to \textsc{madness} AE has reduced to around 1~meV.

\begin{table*}[!h]
\centering
\caption{\label{tab:cysteine_eigen}Comparison of select eigenvalues (eV), for the cysteine molecule using the AE and PSP approaches in MADNESS, PSP implementations in \textsc{BigDFT} and CASTEP, various Gaussian basis sets using the AE approach and the Stuttgart relativistic large core (SRLC) ECP using \textsc{NWChem}, as described in the text.} 
\begin{tabular}{l cc c c ccccc}
 & \multicolumn{2}{c}{\textsc{madness}} & \textsc{BigDFT} & \textsc{castep} &  \multicolumn{5}{c}{\textsc{NWChem}} \\
 & AE & PSP & PSP & PSP & \multicolumn{4}{c}{AE} & ECP \\
 &  &  & &  & aVDZ & aVTZ & aVQZ & aV5Z & SRLC\\
\hline\\[-2.0ex]
HOMO-3 & -7.8867 &		-7.8915 &	-7.8916	 & -7.8619	& -7.8750	& -7.8804 &	-7.8859 &	-7.8859	 & -7.8287\\
HOMO-2 & -6.9024 & 		-6.9099 &	-6.9095 &	-6.8869 &	-6.8654 &	-6.8954 &	-6.9008	& -6.9008	 & -6.9933\\
HOMO-1 & -6.0708 &		-6.0652 &	-6.0652 &	-6.0436	& -6.0355	& -6.0654 &	-6.0709 &	-6.0709 &	-5.9838\\
HOMO & -5.7905 &		-5.7812	& -5.7813	& -5.7503	& -5.7688	& -5.7933 &	-5.7933 &	-5.7906 &	-5.7715\\
LUMO & -1.7365 &		-1.7263 &	-1.7263 &	-1.7128 &	-1.7116	 & -1.7252 &	-1.7334 &	-1.7361 &	-1.8232\\
\end{tabular}
\end{table*}

\begin{table*}[!h]
\centering
\caption{\label{tab:cysteine_core}Core energies and energy splittings between states (eV) calculated using the AE and appropriate mixed approach as defined in Table~\ref{tab:cysteine} as well as various Gaussian basis sets. The atom labels are those indicated in Fig.~\ref{fig:cysteine_mol}.  For the S~2p energy levels there is significant mixing between the states, particularly between $y$ and $z$.  The specific orbital has therefore not been labelled. }
\begin{tabular}{ll cc cccc}
 & & \multicolumn{2}{c}{\textsc{madness}} &  \multicolumn{4}{c}{\textsc{NWChem}}\\
Atom & State & AE & Mixed & aVDZ & aVTZ & aVQZ & aV5Z\\
\hline\\[-2.0ex]
\multirow{5}{*}{S} & 1s	&	-2387.00 &	-2387.00 & -2387.64 & -2387.33 & -2386.65 & -2386.75 \\[0.5ex]
 & 2s	&	-207.82 &	-207.82 &  -208.16 & -208.00	& -207.85 &	-207.82\\[0.5ex]
 & \multirow{3}{*}{2p}	&	-154.91 &	-154.91 & -155.31 &	-155.13 &	-154.95 &	-154.91 \\
 & &	-154.79 &	-154.79 & -155.18	& -154.80 & -154.80 & -154.80\\
 & & -154.61 &	-154.61 & -154.97	& -154.82 & -154.65 & -154.61\\
\hline\\[-2.0ex]
C1 & 1s &		-270.02 &	-270.04 & -270.28	& -270.03 & -270.03 & -270.03 \\[0.5ex]
C2 & 1s &		-267.86 &	-267.87 & -268.13 &	-267.86 &	-267.86	& -267.86\\[0.5ex]
C3 & 1s &		-267.21 &	-267.21 & -267.53	& -267.23 &	-267.22 &	-267.22 \\[0.5ex]
C3 - C1 & 1s & 2.16 &	2.17 &	2.15 &	2.17 &	2.17 &	2.17\\[0.5ex]
C3 - C2 & 1s & 2.81 &	2.83	& 2.75 &	2.80 &	2.81 &	2.81\\[0.5ex]
\hline\\[-2.0ex]
N & 1s &		-377.15 &	-377.16 & -377.63	& -377.18 & -377.16	& -377.15\\[0.5ex]
\hline\\[-2.0ex]
O1 & 1s &		-508.08 &	-508.09 & -508.73	& -508.14 &	-508.10 & -508.09\\[0.5ex]
O2 & 1s &		-506.33 &	-506.33 & -506.99 &	-506.38 &	-506.34	& -506.32 \\[0.5ex]
O2 - O1 & 1s & 1.76 &	1.76 &	1.75 &	1.76 &	1.76 &	1.76\\[0.5ex]
\end{tabular}
\end{table*}

\begin{figure}[!h]
\centering
{\includegraphics[height=0.5\textwidth,angle=-90]{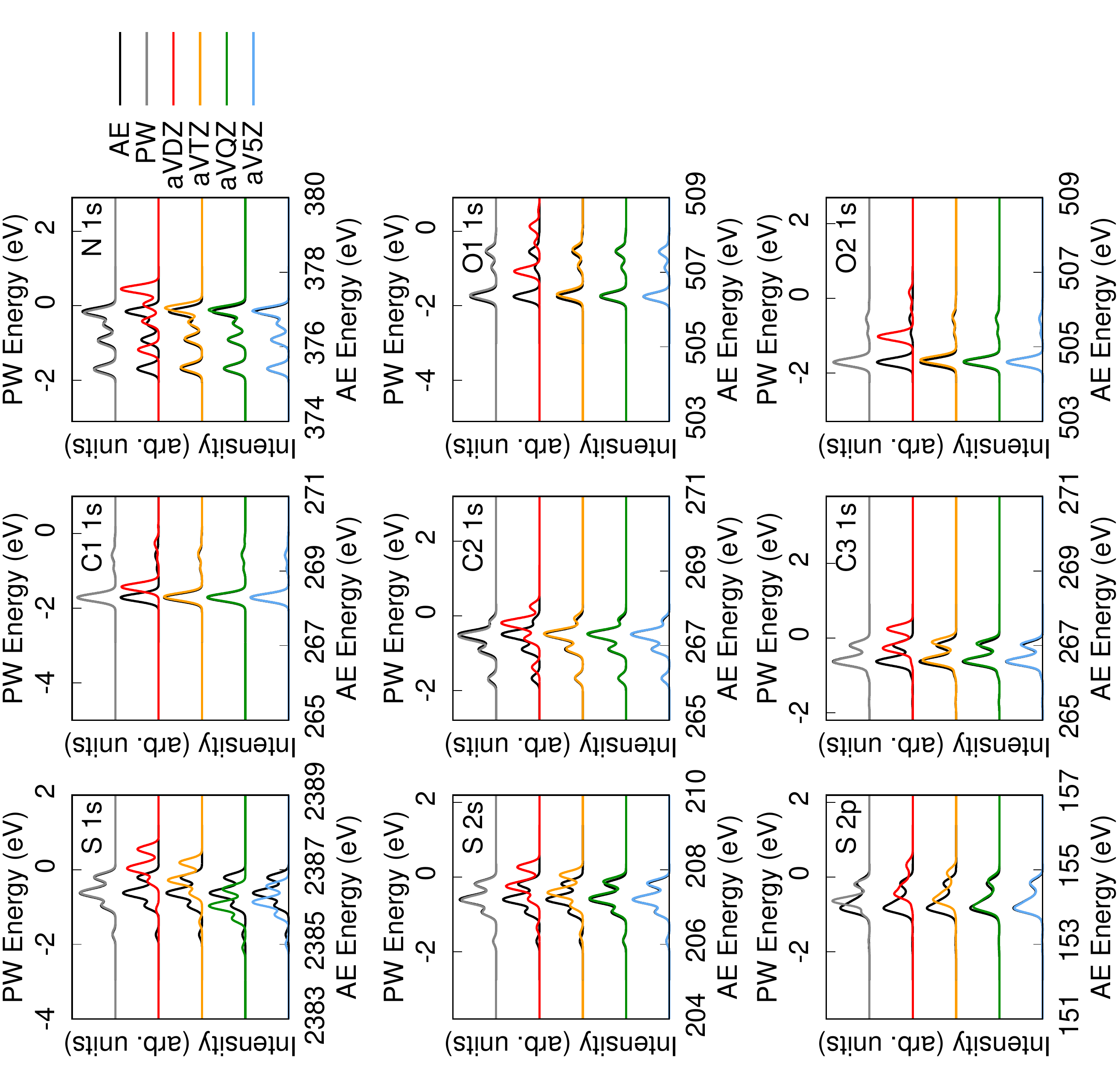}}
\caption{ELNES spectra for the different atoms of the cysteine molecule as labelled in Fig.~\ref{fig:cysteine_mol}, using the AE approach in \textsc{madness}, the PW approach in \textsc{castep} and various Gaussian basis sets in \textsc{NWChem}. Gaussian smearing of 0.1~eV was applied and the intensities were scaled for easier comparison.}
\label{fig:cysteine_eels_atoms}
\end{figure}

We also compare different approaches to ELNES calculations.
This includes those obtained using Gaussian basis sets in \textsc{NWChem} and using the PW code \textsc{castep}. 
As with moldft, the dipole approximation is used in both \textsc{castep} and \textsc{NWChem} to generate the transition matrix elements.  Since we are interested in comparing like-for-like, we do not introduce a core hole into any of the calculations.
Fig.~\ref{fig:cysteine_eels_atoms} shows the ELNES spectra for transitions from each atom in the system.  Since the core states are not accessible in the PW approach, the spectra have been manually aligned, while in all other cases the transition energies are explicitly calculated.  There is excellent agreement between the PW results and \textsc{madness} for all cases except the S~2p spectrum, which is due to the splitting in the core energy levels which cannot be captured with the PW approach.  The Gaussian results converge rather slowly with respect to basis size, especially in the case of S.  However, the shape is already similar for small basis sets.  Furthermore, a closer examination of the core energies, which are given in Table~\ref{tab:cysteine_core}, shows that while the core energies themselves converge slowly, the splitting between core levels converges quickly.

In order to test the mixed scheme, we next investigated several scenarios, which are listed in Table~\ref{tab:cysteine}.  In each case we calculated the occupied and negative energy unoccupied KS states. The error in the eigenvalues compared to the AE values is of the same order of magnitude as for the PSP calculation -- around 10~meV for the set of states considered in Table~\ref{tab:cysteine_eigen}.  In other words, the mixed formalism works as anticipated, with the quality of results only being limited by the quality of the PSP.
In each of the mixed cases, the total wall time was reduced by around a factor of two compared to pure AE. However, no particular effort was made to optimize the implementation of PSPs in \textsc{madness}.  It is therefore likely that further decreases in computational cost could be achieved in the future, e.g.\ by optimizing the calculation of the PSP projectors in the multi-wavelet basis.

\begin{table}[!h]
\centering
\caption{\label{tab:cysteine}The different setups for the cysteine molecule, where \good\ (\bad) denotes the treatment of all atoms of that species at the AE (PSP) level. The column denoted ``ELNES'' indicates whether the calculation of an ELNES spectra is possible.}  
\begin{tabular}{l c ccccc c c}
 && H & C & N & O & S && ELNES \\
\hline\\[-2.0ex]
AE  && \good & \good & \good & \good & \good && \good \\
PSP && \bad & \bad & \bad & \bad & \bad && \bad \\
mixed (H) && \good & \bad & \bad & \bad & \bad && \bad \\
mixed (C) && \bad & \good & \bad & \bad & \bad  && \good  \\
mixed (N) && \bad & \bad & \good & \bad & \bad && \good \\
mixed (O) && \bad & \bad & \bad &\good & \bad && \good  \\
mixed (S) && \bad & \bad & \bad & \bad & \good && \good \\
\end{tabular}
\end{table}

\begin{figure}[!h]
\centering
{\includegraphics[height=0.49\textwidth,angle=-90]{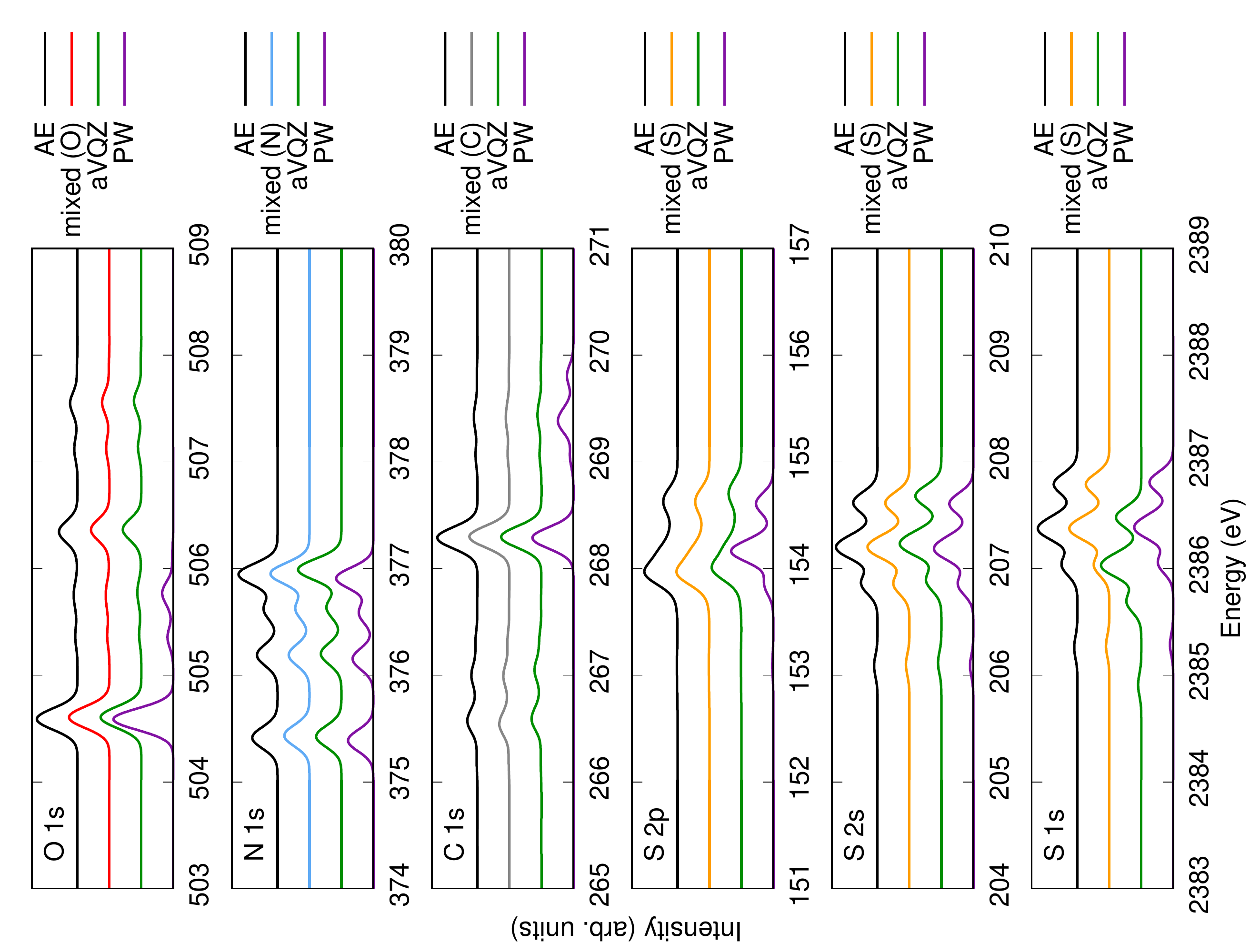}}
\caption{ELNES spectra for the cysteine molecule, for the AE approach and various mixed setups (defined in Table~\ref{tab:cysteine}) in \textsc{madness}, as well as using the PW approach in \textsc{castep} and the AE approach from \textsc{NWChem} using the aVQZ basis set.  The core state is indicated for each plot, and where more than one atom of a given species is present, the average spectrum is shown.
The PW spectra have been manually aligned and intensities have been scaled to facilitate comparison.
Gaussian smearing of 0.1~eV was applied.}
\label{fig:cysteine_eels2}
\end{figure}

Finally, we also calculate ELNES spectra using the mixed approach.
In this case, the spectra can only be calculated where we have access to at least one core state, which is only possible when at least one atom was treated at the AE level. Table~\ref{tab:cysteine} indicates which of the calculation setups were used, while Fig.~\ref{fig:cysteine_eels2} shows the respective spectra, which are also compared with PW and aVQZ results.
Each curve corresponds to the average of all transitions from atoms of that species. 
As can be seen, the PW results show noticeable differences for the C and O 1s spectra, for the same reasons discussed above for the S 2p spectrum.  This is not surprising given that the splittings between core states are of the order of 2~eV.
On the other hand, the mixed and AE results are virtually identical at the applied smearing level, as are the Gaussian results.  The only exception is that the Gaussian results differ for the S 1s spectrum, which in any case would not be measured experimentally.
Indeed, the absolute core energy levels calculated using the different mixed approaches, (given in Table~\ref{tab:cysteine_core}) show excellent agreement with the pure AE results, with the induced error on the order of 0.01~eV.
Thus, should one be interested for example in only the core edge for C, one could easily treat all other atoms at the PSP level at a reduced computational cost, without noticeable loss of accuracy, and indeed with an accuracy comparable to that of a very large Gaussian basis set.

\subsection{Carbon Nanotube}

\begin{figure}[!h]
\centering
{\includegraphics[height=0.3\textwidth,angle=90]{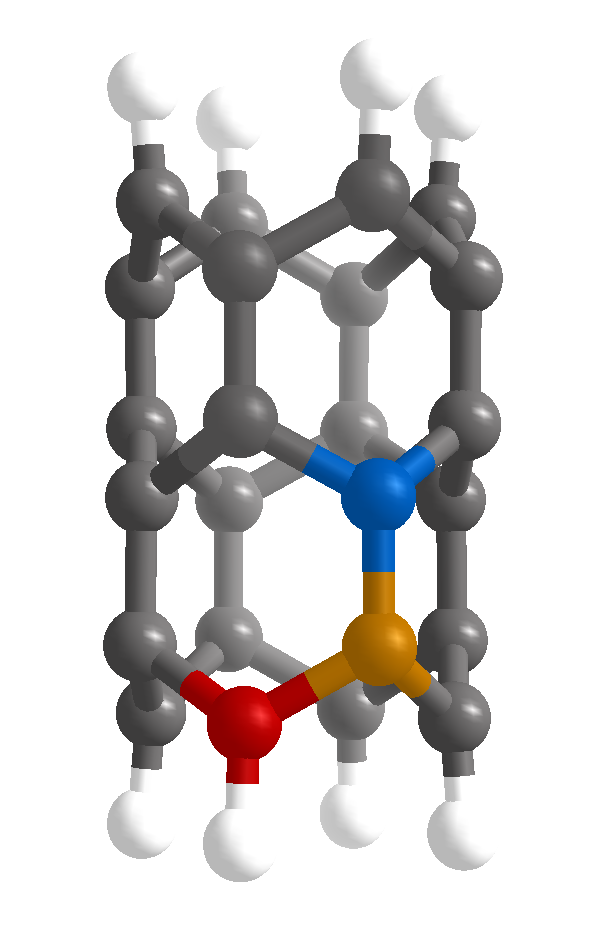}}
\caption{Atomic structure of the SWCNT, with H atoms in white, generic C atoms treated at the PSP level in grey and the atoms with different environments which are treated at the AE level and labelled A, B and C highlighted in red, yellow and blue respectively.}
\label{fig:cnt}
\end{figure}

As our second example, we take a short, hydrogen-terminated (4,0) single walled carbon nanotube, depicted in Fig.~\ref{fig:cnt}.  For this system we are interested in calculating the full ELNES spectrum, however since there are only three C atoms which could be considered to be distinct (i.e.\ different chemical environment, unrelated by symmetry), we might hope to calculate the core excitations from only one atom of each type, and use this to reconstruct the total spectrum.  This can be achieved using a mixed calculation with only three atoms treated at the AE level and the remainder at the PSP level. The AE atoms are labelled A, B and C and indicated in Fig.~\ref{fig:cnt}. 
For the mixed moldft calculation, it was not necessary to explicitly impose any localization on any of the core states, since they remained disentangled.
As well as comparing the spectra originating from the different atoms, we are also interested in the core energies of atoms A, B and C, i.e.\ to what extent there is splitting in the energy levels due to the differing bonding environments. 
As with cysteine, we also compare results with PW and Gaussian basis set approaches.

\begin{figure}[!h]
\centering
\includegraphics[height=0.49\textwidth,angle=-90]{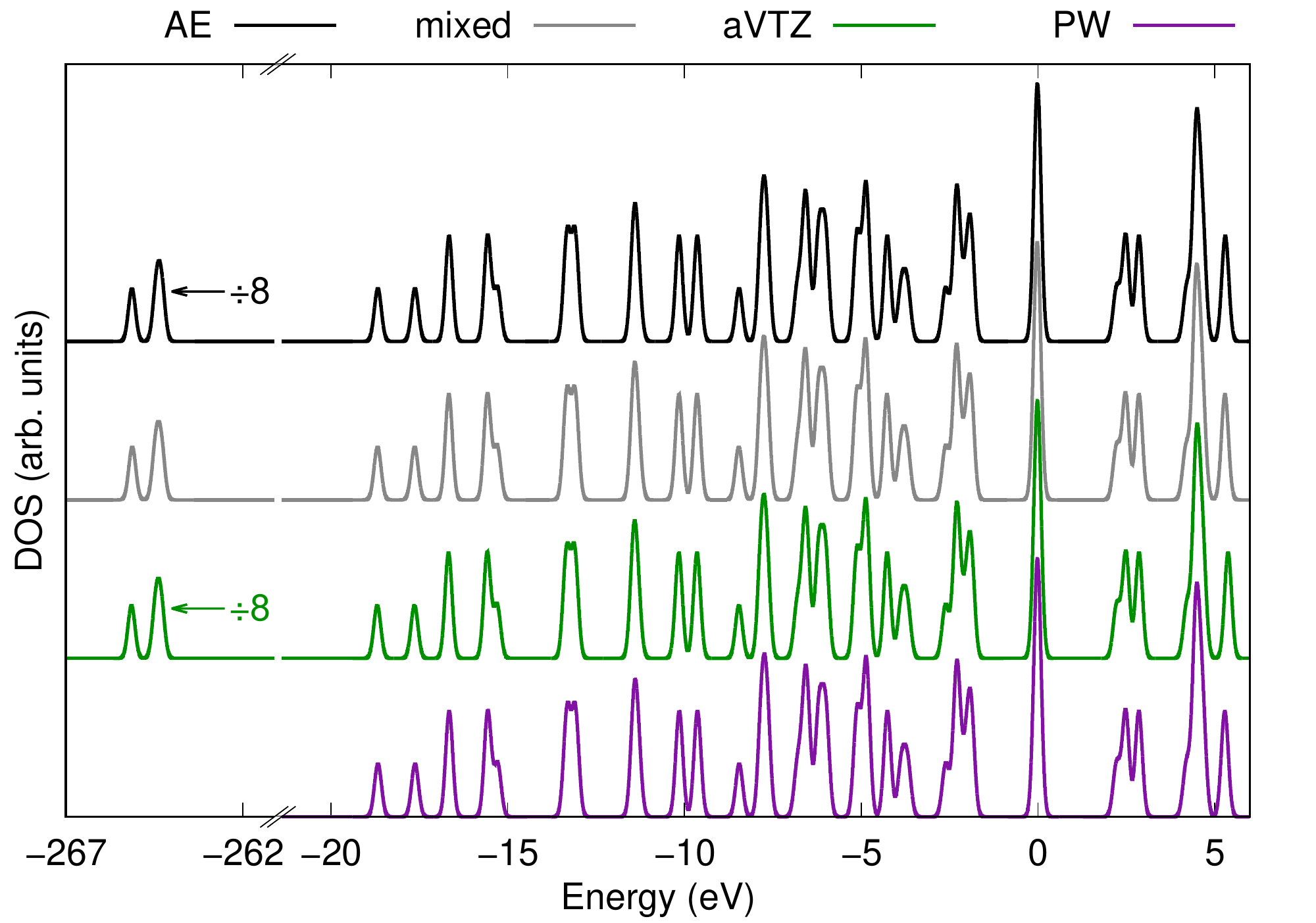}
\caption{DOS for a SWCNT, calculated using the AE and mixed setup of moldft as described in the text, with a PW approach in \textsc{castep} and with a Gaussian basis set in \textsc{NWChem}. Gaussian smearing of 0.1~eV has been applied, while the energies have been shifted so that the HOMO of each setup is at zero. The core energies for the AE and Gaussian calculations (indicated by the arrow in the figure) have been divided by eight to facilitate comparison with the mixed mode DOS.}\label{fig:cnt_dos}
\end{figure}

In the first instance we calculated the density of states (DOS) for the different approaches, shown in Fig.~\ref{fig:cnt_dos}.  As can be seen, the four curves are virtually identical within the level of smearing for the valence and conduction states.  Furthermore, despite the varying setups (different basis sets, type of PSP if used, boundary conditions etc.), the calculated HOMO-LUMO band gaps differ by at most a few meV.  We can therefore be confident that the calculated electronic structure of the different calculations is in agreement, and thus any discrepancies in the calculated ELNES spectra result from the differing approaches to calculating the spectra, rather than as a result of more fundamental differences between the codes.

Considering now the core energies, we compare the mixed and AE results.  There are eight atoms each of types A, B and C and thus we have scaled the core AE DOS by $\frac{1}{8}$ for comparison with the mixed DOS.  Aside from the need for scaling, there is excellent agreement between the two setups.
In order to be more quantitative, in Table~\ref{tab:cnt_core} we give the core energies for the different approaches.
There is non-negligible splitting between the core energy levels, with that of atom A around 0.7~eV lower in energy.  All methods are in good agreement for the splittings, although a larger Gaussian basis set is required to also obtain well converged absolute energies.  
As with cysteine, such a level of splitting, which mainly arises due to the edge of the SWCNT, could have a potentially important affect on the ELNES spectrum.

\begin{table}[!h]
\centering
\caption{\label{tab:cnt_core}Core energies and energy splittings (eV) associated with a particular atom type for the SWCNT calculated using the AE and mixed approaches as well as two Gaussian basis sets.
For the AE and Gaussian values, the energies have been averaged over all instances of atoms with the same symmetry.  The atom labels are those indicated in Fig.~\ref{fig:cnt}.} 
\begin{tabular}{l l cc cc}
& \multicolumn{2}{c}{\textsc{madness}} & \multicolumn{2}{c}{\textsc{NWChem}} \\
Atom & AE & Mixed & aVDZ & aVTZ \\ 
\hline\\[-2.0ex]
A & -270.48 &	-270.46 & -270.81 & -270.49 \\[0.5ex]
B & -269.78 &	-269.79 & -270.12 & -269.80 \\[0.5ex] 
C & -269.64 &	-269.64 & -269.96 & -269.65  \\[0.5ex]
B - A & 0.69 &	0.66 &	0.68 &	0.69  \\[0.5ex] 
C - A & 0.84	&	0.82	&	0.84	&	0.84	\\[0.5ex]
\end{tabular}
\end{table}

\begin{figure}[!h]
\centering
\includegraphics[height=0.49\textwidth,angle=-90]{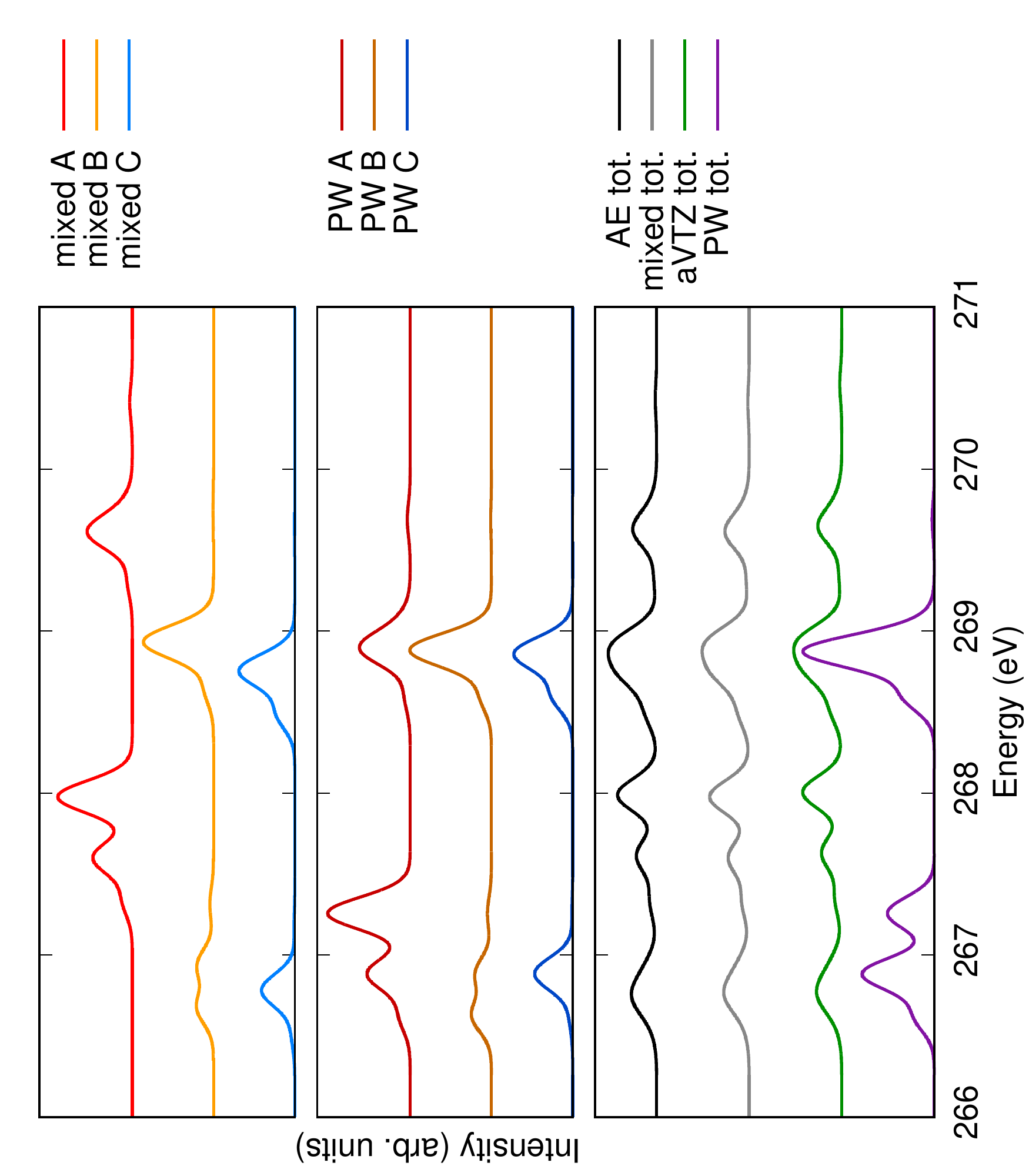}
\caption{ELNES spectra for the C K edge of a SWCNT, calculated in AE and mixed modes in \textsc{madness}, the PW approach in \textsc{castep} and using a Gaussian basis set in \textsc{NWChem}.  In the lower panel the total spectra are plotted (combining the spectra of each atom), while the top two panels show the core edges for atoms A, B and C for the mixed and PW approaches. Gaussian smearing of 0.1~eV has been applied. 
Each of the PW spectra have been manually shifted along the energy axis by a single energy value such that the total spectra can be compared.  Similarly, the intensities have also been scaled to facilitate comparison.}\label{fig:cnt_eels}
\end{figure}

We now turn to the ELNES spectra, which are shown in Fig.~\ref{fig:cnt_eels}.  Alongside the spectra for the three atom types, the total spectra is also shown. In the case of the AE, PW and Gaussian calculations this is merely the sum of the C K edge spectra of each C atom in the system.  For the mixed calculation, we take the three representative spectra and weight them accordingly.  
The AE and mixed spectra are in excellent agreement, demonstrating that our method is able to distinguish between the different carbon atoms, with the relative heights of the different peaks significantly affected by the local environment.  Through judicial choice of atoms treated at the AE level, it is possible to generate the correct averaged spectra without needing to treat all the core states explicitly.  Such an approach could be applied to larger systems which contain atoms which are equivalent by symmetry.
On the other hand, the total PW spectrum shows significant differences with the other approaches.  Upon examining the separate contributions from the three types of atoms, which are plotted in the top three panels of Fig.~\ref{fig:cnt_eels}, it can be seen that the spectra for the different atoms are very similar, with only small differences in the relative peak heights and locations.  As with cysteine, the difference in the total spectra is therefore almost entirely due to the core level splitting which has not been captured in the PW calculation.

\begin{figure}[!h]
\centering
\includegraphics[height=0.49\textwidth,angle=-90]{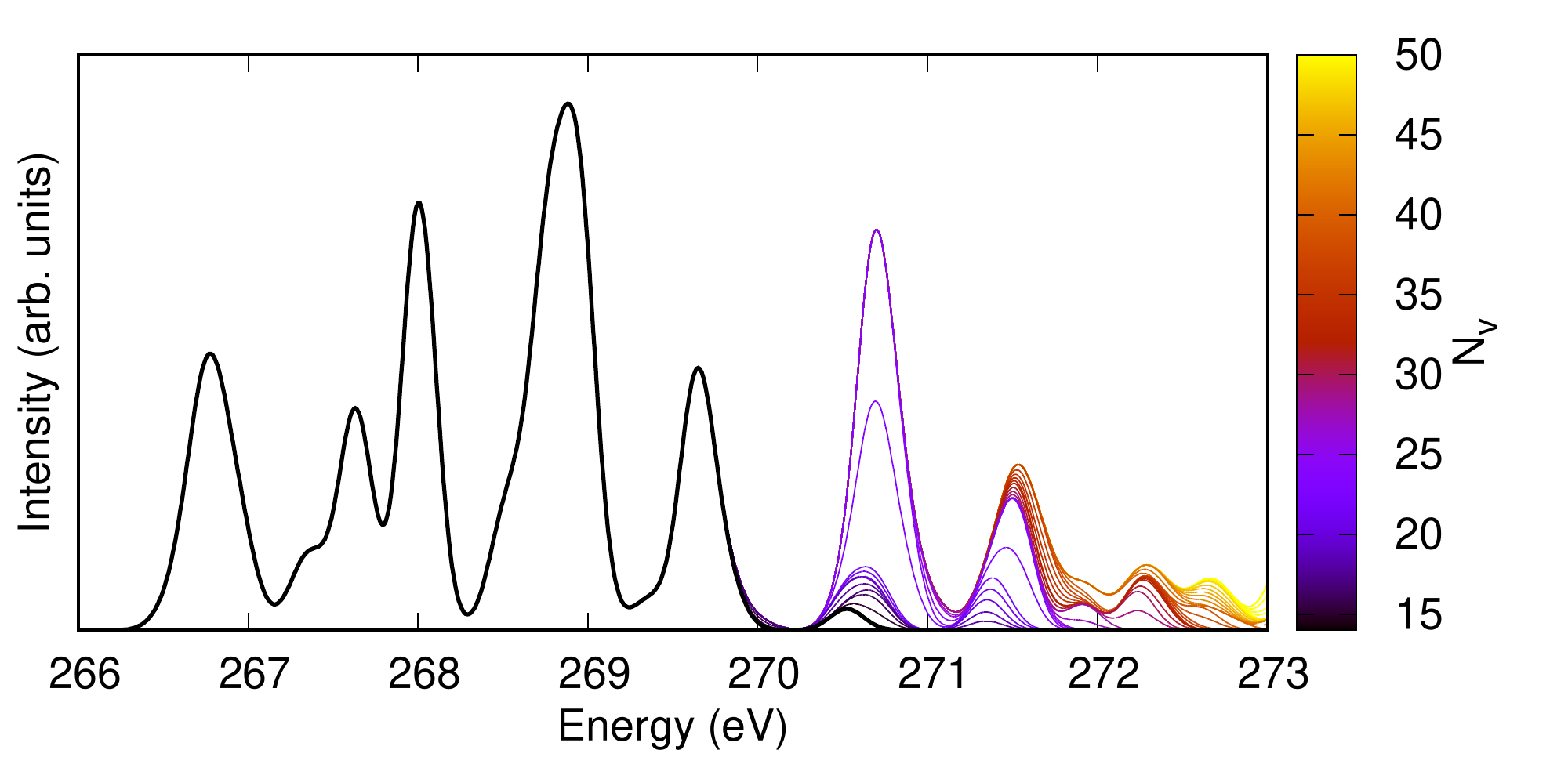}
\caption{ELNES spectra for the C K edge of a SWCNT, calculated using the aVTZ basis set in \textsc{NWChem}. The line colour refers to the number of virtual states, $N_v$, used to generate the spectrum.  The spectrum for $N_v=14$ has been highlighted in thicker black lines. Gaussian smearing of 0.1~eV has been applied.}\label{fig:cnt_eels_empty}
\end{figure}

As previously discussed, only the bound virtual states can be calculated using \textsc{madness}.  Using the aVTZ basis in \textsc{NWChem}, we therefore also explore the convergence with respect to the number of virtual states by calculating ELNES spectra for an increasing number of virtual states, beyond $N_v=14$, which was used in the above.  The results are shown in Fig.~\ref{fig:cnt_eels_empty}.  It can be seen that the spectrum is already converged up to around 270~eV, after which an increasing number of (unbound) virtual states are required.  
We emphasize that the unbound states converge slowly with respect to both the simulation cell size and basis set size~\cite{Boffi2016}, so that the higher energy part of the spectrum requires careful convergence irrespective of the method used.  Nonetheless, provided care is taken to ensure that the number of accessible virtual states is sufficient to converge the spectrum in the region of interest, the restriction to only bound states does not prevent the ability to generate well converged ELNES spectra in a low energy window.

Finally, we note that the level of smearing applied is small compared to typical experimental energy resolutions -- this was chosen to allow for a precise comparison between the different methods.  The impact of core splitting might be less significant for a larger smearing level, however it would still have a noticeable impact on both the position and shape of the peaks. 
Therefore, for systems where there exist atoms of the same species with differing local chemical environments, it is essential to calculate the absolute energy onset of different atoms. 
In the future, it would be interesting to perform a similar comparison taking into account core hole effects, for example in order to assess to what extent the estimation of absolute energy onsets in a PAW approach recovers the values of an AE calculation.  Nonetheless, the mixed AE/PSP approach presented in this work might offer an appealing alternative to both PAW and pure AE approaches, given the possibility of accessing the core states without the full cost of an AE calculation.

\section{Conclusion}

In this paper we have introduced the implementation of norm-conserving HGH-GTH pseudopentials in the multi-wavelet \textsc{madness} molecular DFT code. 
We have validated our implementation through comparison with both PSP and AE calculations using other basis sets. 
The difference between the KS energies calculated using the AE and PSP approaches in \textsc{madness} are shown to be well below the desired accuracy for many applications.
In the first instance, the introduction of such a functionality opens up new possibilities for the treatment of heavy elements or larger molecules in \textsc{madness}.  

We have also presented a mixed AE/PSP approach, wherein atoms within a single calculation may be treated at different levels of theory.  The multiresolution approach of \textsc{madness} automatically refines the basis in areas requiring a greater resolution, thereby achieving a balance between the competing requirements of accuracy and computational efficiency, irrespective of the choice of PSP or AE.   Crucially, this does not require any input from the user.

The availability of a mixed approach is particularly useful in cases where one requires a high degree of accuracy in a particular subset of a system.  For example, for the simulation of a molecule on a surface, where one might choose to treat the molecule and upper layer(s) of the surface at the AE level, and lower layers of the surface at a PSP level.  Furthermore, such an approach is particularly suited to the calculation of ELNES spectra, 
where explicit access to the core states is highly desirable, but only required for a subset of a system.

We have implemented the calculation of ELNES in \textsc{madness} employing the dipole approximation and Fermi's golden rule.  Using examples of a small molecule and short finite SWCNT,  we have shown how our method compares with ELNES spectra calculated using both a PW approach wherein PAW is used to generate the matrix elements, and with Gaussian basis sets.
Through these examples we have demonstrated how one might only treat either select atomic species or select (symmetry-unrelated) atoms at an AE level.  This offers a compromise between the high computational cost of AE ELNES calculations and the inability to directly access the core states in PAW based approaches.
In each case we show excellent agreement between pure AE and mixed calculations.  In both systems the different local chemical environments of atoms of the same species resulted in a splitting in the core energies, which needs to be explicitly calculated to correctly generate the ELNES spectra.  A method which has direct access to the core states at a much lower cost than a full AE calculation is highly useful for such a task.

The examples presented suggest future areas of applicability, as well as avenues for further development.  In particular, it would be desirable in future to also incorporate core hole effects, which in the majority of cases significantly improves the agreement between theoretical and experimental ELNES spectra.  Furthermore, the current approach to calculating the virtual states in moldft is not very robust -- in some cases the inclusion of too many virtual states leads to a failure to converge.  In addition, the energy range is limited due to the fact that the unbound (i.e.\ continuum) virtual states are ill-defined in the multi-wavelet basis.  Alternative approaches to calculating virtual states should therefore be explored in future.

There are also other potential areas of application -- the availability of a high precision code which can operate as AE or PSP introduces new opportunities for benchmarking new flavours or parameterizations of PSPs.  
There has been a recent investment within the community in benchmarking different DFT approaches and codes, notably in comparing a wide range of periodic DFT codes~\cite{Lejaeghere2016}, while multi-wavelets have also been used to benchmark AE basis sets for calculations of molecules~\cite{Jensen2017}.
In a similar spirit, one could use \textsc{madness} to separate errors resulting from different basis sets or other code features from those coming from the choice of PSP.

Furthermore, in order to test the accuracy of a particular PSP, one must typically perform benchmarks for a material containing only one atomic species, e.g.\ elemental solids~\cite{Lejaeghere2016, Prandini2018}, since otherwise it is difficult to disentangle errors resulting from the different PSPs.  However, using the mixed approach, one could also consider materials containing more than one atomic species by treating only the element of interest at the PSP level and all other species at the AE level.  This would allow one to disentangle the errors resulting from different PSPs, without additional approximations due to the basis set. The work presented above could therefore represent a powerful tool for future benchmarking endeavours.

\begin{acknowledgement}
This material is based upon work supported by Laboratory Directed Research and Development (LDRD) funding from Argonne National Laboratory, provided by the Director, Office of Science, of the U.S. Department of Energy under Contract No. DE-AC02-06CH11357.
LER acknowledges an EPSRC Early Career Research Fellowship (EP/P033253/1).
An award of computer time was provided by the Innovative and Novel Computational Impact on Theory and Experiment (INCITE) program. This research used resources of the Argonne Leadership Computing Facility, which is a DOE Office of Science User Facility supported under Contract DE-AC02-06CH11357.
Calculations were also performed on the Imperial College High Performance Computing Service.
LER thanks Dr.\ Anna Regoutz for useful discussions.
Simulation data are available from the author on request.

\end{acknowledgement}

\bibliography{psp}

\end{document}